\documentclass[conference]{IEEEtran}
\IEEEoverridecommandlockouts
\usepackage{cite}
\usepackage{amsmath,amssymb,amsfonts}
\usepackage{graphicx}
\usepackage{textcomp}
\usepackage{xcolor}
\usepackage{algorithm}
\usepackage{algorithmic}
\usepackage[normalem]{ulem}
\usepackage{tikz}
\usepackage{mathrsfs}
\usepackage{subcaption}
\usepackage{multirow}
\usepackage{adjustbox}
\usepackage{algorithm}
\usepackage{algorithmic}
\definecolor{tealshade}{rgb}{0, 0.588, 0.533}
\newcommand{\red}[1]{\textcolor{black}{#1}}
\usepackage{array}

 \usepackage{pgfplots}
\usepackage[labelformat=simple, labelsep=colon]{subcaption}

\def\BibTeX{{\rm B\kern-.05em{\sc i\kern-.025em b}\kern-.08em
    T\kern-.1667em\lower.7ex\hbox{E}\kern-.125emX}}
    
\begin{document}

\title{ED-DAO: Energy Donation Algorithms based on Decentralized Autonomous Organization}


\author{
    Abdulrezzak Zekiye$^\star$, Ouns Bouachir$^\dag$, \"Oznur \"Ozkasap$^\star$, Moayad Aloqaily$^\ddag$ \vspace{5px} \\
    {$^\star${Ko\c{c} University, Department of Computer Engineering, Istanbul, Turkey}}\\
    {$^\dag$College of Technological Innovation (CTI), Zayed University, UAE}\\
    {$^\ddag$Mohamed bin Zayed University of Artificial Intelligence (MBZUAI), UAE}\\
    Emails:{$^\star$\{azakieh22, oozkasap\}@ku.edu.tr, $^\dag$ouns.bouachir@zu.ac.ae, $^\ddag$moayad.aloqaily@mbzuai.ac.ae
    }
}

\IEEEoverridecommandlockouts

\maketitle
\IEEEpubidadjcol

\begin{abstract}
Energy is a fundamental component of modern life, driving nearly all aspects of daily activities. As such, the inability to access energy when needed is a significant issue that requires innovative solutions. In this paper, we propose ED-DAO, a \red{novel} fully transparent and community-driven decentralized autonomous organization (DAO) designed to facilitate energy donations. We analyze the energy donation process by exploring various approaches and categorizing them based on \red{both} the source of donated energy and funding origins. We propose a novel Hybrid Energy Donation (HED) algorithm, which enables contributions from both external and internal donors. External donations are \red{payments} sourced from entities such as charities and organizations, where energy is sourced from the utility grid and prosumers. Internal donations, on the other hand, come from peer contributors with surplus energy. \textcolor{black}{HED prioritizes donations in the following sequence: peer-sourced energy (P2D), utility-grid-sourced energy (UG2D), and direct energy donations by peers (P2PD).} By merging these donation approaches, the HED algorithm increases the volume of donated energy, providing a more effective means to address energy poverty. \red{Experiments were conducted on a dataset to evaluate the effectiveness of the proposed method. The results showed that HED increased the total donated energy by at least 0.43\% (64 megawatts) compared to the other algorithms (UG2D, P2D, and P2PD).}  

\end{abstract}

\begin{IEEEkeywords}
Energy donation, decentralized autonomous organization, blockchain, microgrids.
\end{IEEEkeywords}

\section{Introduction}
Worldwide energy consumption increased by approximately 48\% between 2000 and 2022 and is projected to grow by an additional 6.7\% by 2050, reaching an estimated 635 exajoules \cite{statista_energy_consumption}. This upward trend is primarily attributed to an increasing reliance on energy for various systems, such as heating and electric vehicles. However, as energy demand rises, so does the challenge of ensuring sufficient supply, thus exacerbate the prevalence of energy poverty. Energy poverty refers to the lack of access to essential energy services required for economic resilience and human development \cite{reddy2000energy}. It is a commonly underestimated issue, often mistakenly perceived as a challenge limited to developing countries. Recent statistics indicate that energy-related hardships affected 10.6\% of households globally in 2023, up from 6.9\% in 2021 \cite{eu_energy_poverty}, underscoring its status as a global concern demanding effective solutions. 

Although energy donation has emerged as a potential approach to mitigate energy poverty, it remains underexplored in academic research, with few studies exploring its effectiveness \cite{cali2021novel, khalid2022optimizing}. 
\red{Traditional donation systems rely on a central entity, such as a governmental office, to decide whether any particular consumer deserves donations or not. In the case of energy donation, these central entities may either provide financial assistance directly to consumers or pay a portion of their utility bills on their behalf. A remarkable example of this is the Low Income Home Energy Assistance Program (LIHEAP) \cite{liheap}.}

In \cite{zekiye2024blockchain, zekiye2024analysis, cali2021novel}, blockchain was used to provide a degree of decentralization and transparency. In our work, we propose complete decentralization of the system, reconfiguring it into a decentralized autonomous organization (DAO). DAOs operate over the blockchain, enabling the decentralization of both operations and governance. This structure eliminates centralized decision-making, instead facilitating collective decision-making by members, thereby ensuring a fully distributed governance structure \cite{wang2019decentralized}.




\red{The primary issue addressed by this research is the inefficiency of traditional energy donation systems in alleviating energy poverty. Furthermore, we address the absence of a fully decentralized framework in which governance is controlled by the donors.} The main contributions of this study are as follows:
\begin{enumerate}
\item We propose an Energy Donation-DAO (ED-DAO) framework to enhance decentralization in energy donation processes. In this approach, donation recipients are identified autonomously by members of the decentralized autonomous organization (DAO), ensuring an equitable and decentralized selection process.

\item We analyze and classify various energy donation algorithms and ultimately propose Hybrid Energy Donation (HED) algorithm. This algorithm allocates energy donations to recipients based on three mechanisms: peer-to-peer community sourcing (P2D), utility grid sourcing (UG2D), and direct peer donations (P2PD). 

\item Our experimental evaluations compare these donation scenarios: peer-sourced energy, utility-grid-sourced energy, peer donations, and our proposed HED model. This model prioritizes donations in the order of peer-sourced energy, followed by utility-grid-sourced energy, and lastly, direct peer donations.

\item Key findings from the experiments indicate that the HED model significantly increases total donated energy  compared to the P2D method. Additionally, the HED model \red{increased the donated energy by 0.43\%, compared to the P2PD approach. This 0.43\% increase is equivalent to 64 megawatts, where 1 megawatt can power up approximately 30 homes a day \cite{EIA_Electricity_Use_Homes}}. In scenarios where peer donations are limited or absent, the P2D approach proves more effective than UG2D, doubling the amount of donated energy.

\end{enumerate} 

This paper is organized as follows: Section \ref{rw} discusses related works. Section \ref{eddao} presents the usage of DAO as the main system for energy donation. Different energy donation scenarios, along with our proposed hybrid energy donation algorithm, are presented in Section \ref{energy-donation}. Sections \ref{experiments} and \ref{discussion} include the experimental results and the discussions. Finally, \ref{conc} concludes the paper.

\section{Related Works} \label{rw}

In this section, we discuss the related works considering two key aspects: (1) mitigating energy poverty as in \cite{cali2021novel, khalid2022optimizing, zekiye2024blockchain, zekiye2024analysis}, and (2) utilizing DAOs in the energy field \cite{ding2024customized, sharma2023energydao, aoun2024bottom}.

In \cite{zekiye2024blockchain, zekiye2024analysis}, energy sharing mechanisms have been proposed to mitigate energy poverty, where sharing refers to transacting energy for non-monetary purposes\red{, such as futuristic or indirect payments (not by the receiving donee).} Three sharing algorithms, namely centralized sharing, P2P sharing, and selfish sharing, were proposed and evaluated by simulations. \red{In those sharing algorithms, prosumers can temporarily access energy receiver peers' batteries as compensation for the energy they provide.}.

In \cite{cali2021novel}, a Distributed Ledger Technology (DLT)-based solution to enable anonymous energy donation has been proposed. In this approach, donation recipients are identified in a centralized manner through a governmental social department. Based on the source of the donations, two different donation mechanisms were identified: internal donations and external donations. In external donations, donors are from outside the community, such as charities. In internal donations, the donors are from within the community, where they voluntarily and equivalently cover the energy cost. 

A framework for energy donation in smart grids during crises was proposed in \cite{khalid2022optimizing}. In this framework, microgrids transact their surplus energy either for charitable reasons or to get future benefits. Energy donors would be rewarded in different forms, such as reputation scores or prioritized services. Moreover, energy donors would be compensated by the utility grid once the service is restored.

The utilization of Decentralized Autonomous Organizations in the energy field is noted in \cite{ding2024customized, sharma2023energydao,aoun2024bottom}. In \cite{ding2024customized}, the use of DAOs was proposed to serve as an energy management platform for smart buildings. In their solution, DAO would enable peers to decide on the operational rules of the donation process. More specifically, the aim of the proposed DAO is enhancing the participation rate of buildings in the peer-to-peer day-ahead energy trading.

Converting the centralized decision making in energy communities to a decentralized form was proposed in \cite{sharma2023energydao}. The proposed concept, EnergyDAO, would enable peers to own energy producing and storing facilities, and have a role in shaping the governing rules. 
%
%
Inspired by DAOs, authors in \cite{aoun2024bottom} proposed decentralized autonomous substations (DAS). DAS fully decentralizes the functionality of electrical grid's centralized control centers (CCC). By utilizing smart contracts as a decentralized data management and control \red{platform}, the consequences that result from cyberattacks on CCCs or service failures are mitigated. Additionally, based on the collected data, DASs can make autonomous decisions that lead to: (1) enhancing the grid functionalities, (2) enhancing energy transactions and production, and (3) improving the management and response to natural disasters.

In the light of the research in \cite{cali2021novel}, the donation algorithms we present, P2D and UG2D, could be considered as external donation approaches. In contrast, DP2P follows an external approach. Finally, HED could be considered a hybrid approach that merges external and internal donations to increase the donated energy amount. External donations in the HED algorithm are represented by the external donations paid to the grid or prosumers to send energy to peers in need, and internal donations are represented by the energy donated by peers directly to other peers, without receiving any form of payment. 

Table \ref{tbl:comp} highlights the contributions of this paper, where it is noticeable that our paper offers a novel hybrid energy donation algorithm, in addition to maximizing decentralization by utilizing DAOs.

\begin{table*}[t] 
\centering
\caption{Comparative analysis of ED-DAO and related works} 
\renewcommand{\arraystretch}{1.5} 
\begin{tabular}{|l|c|c|c|l|} 
\hline
\textbf{Ref.} & \parbox[c]{0.8cm}{\textbf{Energy Poverty}}  &  \parbox[c]{1.5cm}{\textbf{Decentralized Operations}} & \parbox[c]{1.5cm}{\textbf{Decentralized Governance}} &    \textbf{Solution}  \\
\hline 
\textbf{\parbox[c]{4cm}{Energy Management DAO \cite{ding2024customized}} } & X &  $\checkmark$ &  $\checkmark$ & DAO for energy management in smart buildings.  \\
\hline 
\textbf{EnergyDAO \cite{sharma2023energydao}} & X & $\checkmark$ & $\checkmark$ & Conceptual framework for decentralizing energy communities. \\

\hline 
\textbf{DAS \cite{aoun2024bottom}} & X & $\checkmark$ & $\checkmark$ & DAS for decentralizing electrical grid's centralized control centers. \\
\hline 
\textbf{\parbox[c]{4.5cm}{Energy donation in smart grids \cite{khalid2022optimizing}}} & $\checkmark$ & X & X & Addressed energy poverty in crises use case. \\
\hline 
\textbf{\parbox[c]{4.5cm}{DLT-based Energy Donation \cite{cali2021novel}}} & $\checkmark$ & $\checkmark$ & X & Blockchain for decentralization. Internal or external donations. \\
\hline 
\textbf{ED-DAO} & $\checkmark$ & $\checkmark$ & $\checkmark$ & DAO for decentralization. Internal and external energy donation. \\

\hline 
\end{tabular} 
\label{tbl:comp} 
\end{table*}

\section{ED-DAO System Model}\label{eddao}
\begin{figure}
    \centering
 \includegraphics[width=0.6\linewidth]{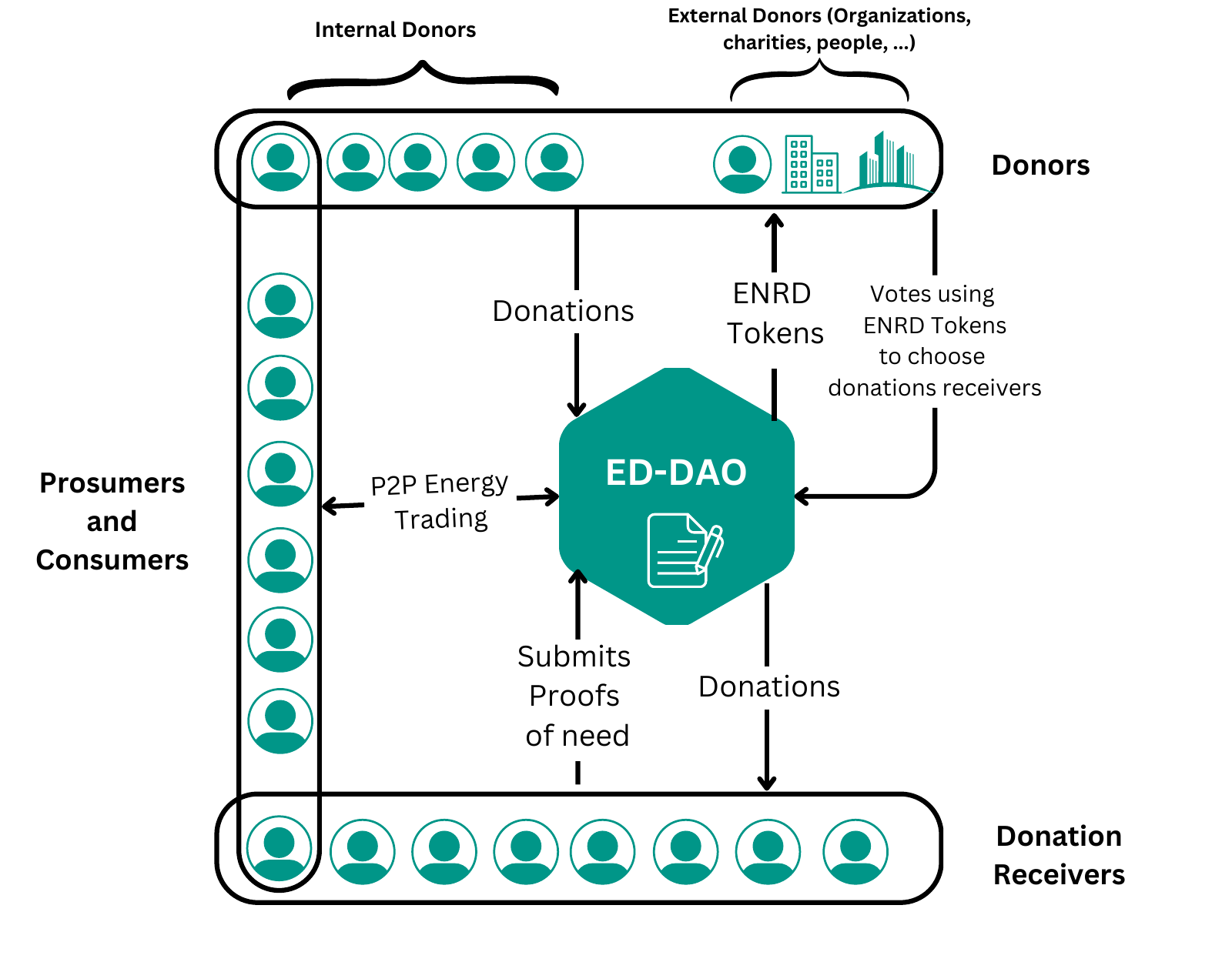}
        \caption{ED-DAO Overview where donors decide who can receive donations by voting using ENRD tokens}
        \label{fig:ED-DAO_overview}
        
    \vspace{-5mm}
\end{figure}

Our proposed ED-DAO system is illustrated in Fig \ref{fig:ED-DAO_overview}, where the main entities are consumers, prosumers, donors, and donation receivers. Prosumers are peers with their own energy resources, such as solar panels, where they might have excess energy at some time. Consumers are peers that do not have their own energy resources. There are two types of donors: money donors and energy donors. In our system, money donors are from outside the system, thus we call them external donors. Those donors can be, but are not limited to, charitable entities and governmental offices. Energy donors are the peers who donate energy to other peers. Donation receivers are peers that have energy need with no funds to acquire it. In traditional systems, donation receivers are identified and verified by some centralized entities. In our proposed ED-DAO, donation receivers are identified in a decentralized manner and through voting by the members of the community. 


\subsection{Governance}
DAOs transform the governance control from a centralized approach to a decentralized one. In ED-DAO, the governance includes making any decision by voting. One of the important things to vote on is who is eligible to receive donations and who is not. To enhance decentralization within the ED-DAO and prevent any individual from gaining excessive influence, the voting process is designed to require voters to burn tokens. This means that each time a person votes, they permanently lose some of their tokens, which gradually reduces their voting power over time. This approach limits any single person’s ability to dominate the decision-making process, promoting a fairer distribution of influence within the ED-DAO.

\subsection{Tokenization}
Tokens are an important part of DAOs since they grant members the right to participate in the governance process. Tokenization refers to the process of creating and issuing tokens that represent an asset, right, or value on a blockchain. Members of a DAO use the tokens to vote. Thus, in order to keep the DAO 
\red{democratized}, it is important to well-define the tokenization policy. EnergyDonor (ENRD) is the token that is used in our proposed ED-DAO. ENRD tokens are minted only when energy is donated, and are sent to the donor. 
\red{Generally, in DAOs, different voting models exist, such as Token-based voting and Quadratic voting \cite{han2025review}, where the tokens used for voting could be redistributed, or burnt (destroyed).} In addition to the usage of ENRD tokens in the voting process, the tokens themselves can be seen as a proof-of-donation, 
\red{as they are exclusively} issued to donors. \red{This functionality may be necessary for regulatory compliance or tax-related purposes.}  

\subsection{Energy Exchange}
Energy is exchanged through ED-DAO by means of two different approaches, trading and donating. ED-DAO facilitates energy trading first between prosumers and consumers, then it mediates energy donation. Pricing is controlled by ED-DAO through an equation that satisfies both buyers and sellers. 
The pricing equation is provided in \eqref{eq:pricing}, where 
\(\text{p}_t\) represents the calculated price at the current time step \(t\). \(\text{up}_t\) denotes the utility price at time step \(t\), and \(R_t\), and \(O_t\) corresponds to the total trading requests submitted by consumers and the total offers made by prosumers at this time step, respectively. In other words, \(R_t\) reflects the demand side activity, while \(O_t\) indicates the supply side contributions, each influencing the dynamics of the calculated price.

\begin{equation}
\begin{aligned}
\text{p}_t = \max(FiT, \min(\text{up}_t, &  \frac{R_t}{\text{O}_t} \cdot \frac{\text{p}_{t - 1} + \text{p}_{t - 2} + \text{p}_{t - 3}}{3}))
\end{aligned}
\label{eq:pricing}
\end{equation}

\red{This pricing equation determines a market-clearing price that remains below the utility grid price, ensuring consumer cost savings, while exceeding the feed-in tariff (FiT), thereby maximizing seller revenue. In addition, the price changes according to the demand-to-supply ratio.} \red{Although alternative pricing methods could be employed in the proposed system, it is crucial to ensure that the energy price between peers remains lower than the utility grid price in order to maintain the effectiveness of the proposed approaches. Otherwise, obtaining energy directly from the grid would present a more reasonable and cost-effective solution.}

\subsubsection{Energy Trading}
ED-DAO matches between energy offers submitted by prosumers and energy requests submitted by consumers, on a first-come first-served manner.
\subsubsection{Energy Donation}
This part of ED-DAO is responsible for handling donation requests by peers in need. In Section \ref{energy-donation}, we identify and propose several different methods to handle donation requests in ED-DAO.  

\section{Energy Donations}\label{energy-donation}

\begin{figure}
    \centering
    \includegraphics[width=0.74\linewidth]{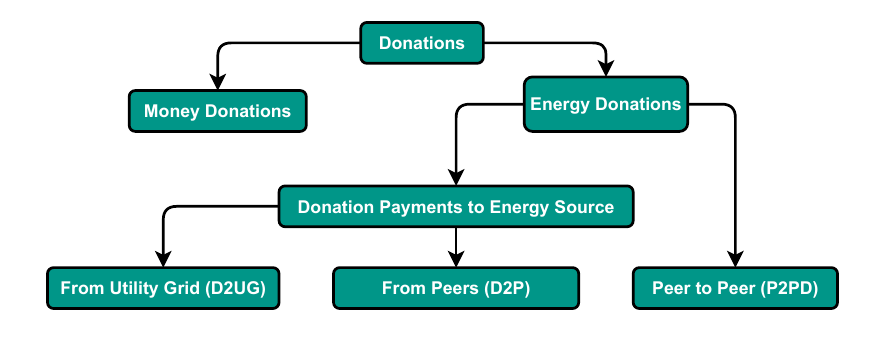}
    \caption{Overview of different donation scenarios}
    \label{fig:donation-scenarios}
    
    \vspace{-5mm}
\end{figure}

Donating to people in need is an important way to mitigate energy poverty. Figure \ref{fig:donation-scenarios} shows different scenarios that are possible in the donation process. Depending on what donation receivers are acquiring, we can categorize the donations into two types: money donations and energy donations.

\subsection{Money Donation}
In this scenario, donations are received by the donation organizer, ED-DAO in our proposal. In case a consumer needs energy without having funds, he will ask for energy to be donated, and the ED-DAO will send the consumer money to buy energy.

\subsection{Donation Payments to Energy Source}\label{D2ES}
In this scenario, donations are received by the donation organizer. In case a consumer needs energy without having funds, he will ask energy to be donated. If the donation organizer has received funds, it will buy energy, and this energy will be transacted directly to the donation receiver. 
Energy can be sourced from the utility grid or prosumers. Based on that, we can identify two donation scenarios.
\subsubsection{Donations Payments to Utility Grid (UG2D)} In this case, ED-DAO pays the utility grid to get the energy sent to the donation receivers.\label{UG2D}
\subsubsection{Donations Payments to Prosumers (P2D)} ED-DAO 
 in this case will pay prosumers who have excess energy to send this excess to donation receivers. \label{P2D}



\subsection{Peer-to-peer Donations (P2PD)}\label{P2PD}
In this scenario, the energy is donated directly by the peers with extra energy to the peers in need, without involvement of any form of payment. Although the donations are by peers, and to peers, a donation manager is still needed to match the donors with donation receivers, where the donation manager in our system is fully decentralized by utilizing ED-DAO.

\subsection{Hybrid Energy Donation (HED)}\label{HED}
\begin{figure}
    \centering
    \includegraphics[width=0.6\linewidth]{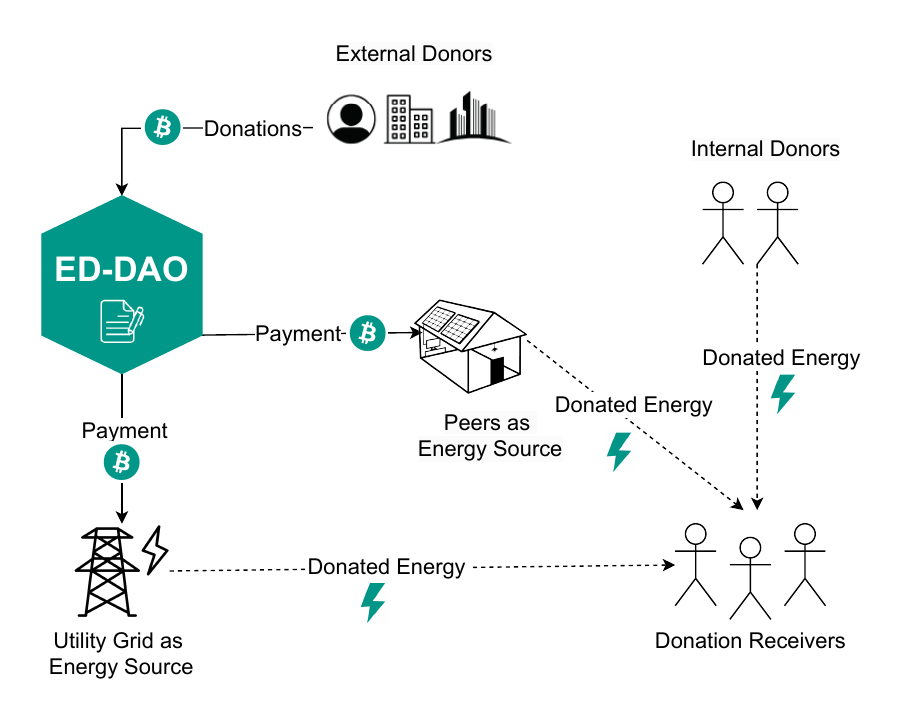}
    \caption{Overview of our proposed HED mechanism}
    \label{fig:hed}
\end{figure}

In this paper, we propose a hybrid energy donation scenario in which energy is donated by sourcing it from the utility grid or peers, or donated directly by the peers, as shown in Algorithm \ref{alg:hed}, \red{where Table \ref{tab:symbol_definitions} illustrates the meanings of the symbols}. In our proposed HED algorithm, if the ED-DAO has funds from money donations, it will purchase energy from peers and donate it to those in need. If the peers do not have excess energy, then energy will be sourced from the utility grid. In the case of not having sufficient funds, then peers with excess energy would donate energy directly to the peers in need. In other words, this donation scenario consists of Donations Payments to Prosumers (P2D), Donations Payments to Utility Grid (UG2D), and Peer-to-peer Donations (P2PD), as shown in Figure \ref{fig:hed}.

\begin{table}[h!]
\centering
\caption{\red{Symbols and definitions.}}
\label{tab:symbol_definitions}
\begin{tabular}{|c|l|}
\hline
\textbf{Symbol} & \textbf{Definition} \\ \hline
$F_a$          & Funds available in ED-DAO \\ \hline
$P_e$          & Set of peers with excess energy \\ \hline
$P_n$          & Set of peers in need \\ \hline
$C_p$          & Cost to pay a prosumer (peer with excess energy) \\ \hline
$C_u$          & Cost to pay the utility grid \\ \hline
$p$            & Individual prosumer (peer with excess energy) \\ \hline
$r$            & Individual donation receiver (peer in need) \\ \hline
\end{tabular}
\end{table}

\begin{algorithm}
\caption{\red{Hybrid Energy Donation Mechanism (HED)}}
\begin{algorithmic}[1]
    \STATE \textbf{Inputs:} $F_a$, $P_e$, $P_n$
    
    \FOR{$r \in P_n$}
        \FOR{$p \in P_e$}
            \IF{$F_a \geq C_p$}
                \STATE Pay $C_p$ to $p$
                \STATE Donate energy to $r$
                \STATE Deduct $C_p$ from $F_a$
            \ENDIF
        \ENDFOR
    
        \IF{$F_a \geq C_u$}
            \STATE Pay $C_u$ to the utility grid
            \STATE Donate energy to $r$
            \STATE Deduct $C_u$ from $F_a$
        \ENDIF

    \FOR{$p \in P_e$}
        \IF{$p$ is willing to donate energy}
            \STATE Donate energy to $P_n$
        \ENDIF
    \ENDFOR
    
    \ENDFOR
\end{algorithmic}
\label{alg:hed}

\end{algorithm}

\section{Simulations and Results} \label{experiments}
In our simulations, the assumption is that all peers requesting donations are eligible to receive them. Furthermore, any peer with surplus energy after the trading process willingly donates it to those in need. \red{Additionally, we assumed that no peers have energy storage capabilities.}

The dataset that represents energy production and consumption for 69 entities 
in Estonia is used where the data represent hourly production and consumption for each entity spanning over 21 months, between 01-09-2021 and 29-05-2023 \cite{predictEnergyBehaviorOfProsumers}. As in \cite{zekiye2024analysis}, we assumed that each entity is a microgrid since it consists of several consumption points.

To simulate peers' balances, we included a monthly balance for each peer, since the dataset lacks that information. This balance is calculated as a percentage of the amount the peer would pay to the grid\red{, and it would be used to buy energy from other peers in the simulations}. The amount each peer would pay to the grid is based on their energy need \red{from the dataset} multiplied by the energy price. In our experiments, we tested different balance scenarios by setting the percentage to 0.05\%, 0.5\%, and 2\%. Based on the peer's balance in a timestep, the peer can decide to either sell energy, buy energy, or ask for donation. If the peer has excess energy, he would offer it for selling, if he needs energy and 
\red{his balance is sufficient to purchase it}, he would ask to buy energy. If 
\red{the peer's balance is not sufficient to purchase energy from peers}, he would ask energy to be donated. Whether the peer would be a donation receiver or not is decided in a decentralized manner by voting in ED-DAO. 


To simulate the amount of external donations, we assumed that the donations would \textcolor{black}{be} sent by external donors monthly. The total amount of \textcolor{black}{money donated} in a month was estimated as the mean of the monthly added balances. The total amount of externally received donations based on the different \textcolor{black}{percentages of balance} are provided in Table \ref{tab:donations}.

\begin{table}
    \centering
        \caption{Total amount of external donations using different balance percentages.}

    \begin{tabular}{|c|c|c|c|}
    \hline
        \textbf{Balance Percentage} & 0.05\% &  0.5\% &  2\%\\
        \hline
        \textbf{Total Donations (EUR)} & 506 & 5063 & 20253 \\
        \hline
    \end{tabular}
    \label{tab:donations}
\end{table}

All experiments were conducted in Python, where four different energy donation algorithms—UG2D, P2D, P2PD, and HED—across the entire time range are simulated. In each timestep, energy was first traded, followed by energy donation according to the specific donation algorithm.

\subsection{Donated Energy Analysis}

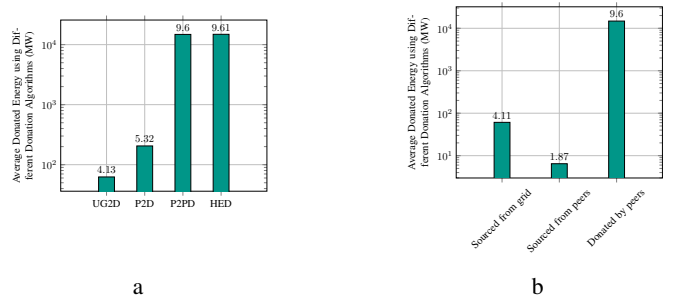
\begin{figure}[h]  
    \centering 
     \begin{subfigure}[b]{0.4\linewidth}
    \centering
        \begin{tikzpicture}[scale=0.4] 
          \begin{axis}[
                ybar,
                symbolic x coords={UG2D, P2D, P2PD, HED},
                xtick=data,
                ylabel={Average Donated Energy using Different Donation Algorithms (MW)},
                ymin=0,
                bar width=15pt,
                enlarge x limits=0.4,
                nodes near coords,
                grid=major,
                ymode=log,
                ylabel style={ align=center, text width=10cm }, 
            ]
            \addplot[fill=tealshade] coordinates {(UG2D, 62.45) (P2D, 205.17) (P2PD, 14.8e3) (HED, 14.864e3)};
            \end{axis}
        \end{tikzpicture}
        \caption{ }
        \label{fig:average-donated-energy}
    \end{subfigure} \hfill
    \begin{subfigure}[b]{0.4\linewidth}
        \begin{tikzpicture}[scale=0.4]
          \begin{axis}[
                ybar,
                symbolic x coords={Sourced from grid, Sourced from peers, Donated by peers},
                xtick=data,
                ylabel={Average Donated Energy using Different Donation Algorithms (MW)},
                ymin=0,
                bar width=15pt,
                enlarge x limits=0.4,
                nodes near coords,
                grid=major,
                ymode=log,        
                xticklabel style={rotate=45}, 
                ylabel style={ align=center, text width=10cm }, 
            ]
            \addplot[fill=tealshade] coordinates {(Sourced from grid, 60.9) (Sourced from peers, 6.46) (Donated by peers, 14.8e3) };
            \end{axis}
        \end{tikzpicture}
        \caption{ }
        \label{fig:hed-source}
    \end{subfigure}
\caption{\red{(a) Average of donated energy across different balance scenarios for UG2D, P2D, P2PD, and HED and (b) Source of average donated energy across different balance scenarios for HED.}}   
    \label{fig:donation-scenarios-combined}
    \end{figure}

Table \ref{tbl:results} shows the amount of donated energy, for each donation algorithm and balance percentage combination. \textit{External Donations (MW)} represents the amount of donated energy in megawatt, using the money donations from external donors, where the energy is sourced either from the utility grid (in the case UG2D algorithm), or from the peers (in the case of P2D algorithm). \textit{Internal Donations (GW)} represents the amount of donated energy by peers in gigawatt. \textit{Total Donated (MW)} is the sum of internal donations and external donations in megawatts. It is notable that the total amount of donated energy in the HED algorithm is superior compared to UG2D, and P2D algorithms, and slightly better than the P2PD. 
Figure \ref{fig:average-donated-energy} shows the average of total donated energy using the balance percentages 0.05\%, 0.5\%, and 2\%. As shown in figure, the amount of donated energy has increased by \red{228}\%, comparing P2D with UG2D. Comparing P2PD and HED to P2D, it is notable that the donated energy amount increased by \red{7113}\%, and \red{7144}\% respectively.
Figure \ref{fig:hed-source} shows the source of the total donated energy in the HED algorithm, where \red{0.40}\%  of the energy was sourced from the utility grid, \red{0.04}\% was sourced from the peers, and \red{99.56}\% was donated by the peers.



\begin{table*}[t] 
\centering 
\caption{\red{Donated energy of different energy donation scenarios using various simulated balances.}} 
\begin{tabular}{|l|cccc|cccc|cccc|} 
\hline
\multirow{2}{*}{\textbf{Metric}} & \multicolumn{4}{c|}{\textbf{Balance Percentage: 0.05}} & \multicolumn{4}{c|}{\textbf{Balance Percentage: 0.5}} & \multicolumn{4}{c|}{\textbf{Balance Percentage: 2}}\\
\cline{2-13}
& \textbf{UG2D} & \textbf{P2D} & \textbf{P2PD} & \textbf{HED} & \textbf{UG2D} & \textbf{P2D} & \textbf{P2PD} & \textbf{HED} & \textbf{UG2D} & \textbf{P2D} & \textbf{P2PD} & \textbf{HED} \\
\hline 
\textbf{External Donations (MW) } & \red{5.17} & \red{8.41} & N.A. & \red{5.17} & \red{31.3} & \red{116.1} & N.A. & \red{34.35} & \red{150.9} & \red{491.0} & N.A. & \red{153.6}\\ 

\hline 
\textbf{Internal Donations (GW) } & N.A. & N.A. & \red{21.6} & \red{21.6} &  N.A.&  N.A. & \red{16.8} & \red{16.8} & N.A.&  N.A.& \red{6.0} & \red{6.0} \\ 
\hline 
\textbf{Total Donated (MW)} & \red{5.17} & \red{8.41} & \red{21.6e3} & \red{21.605e3} & \red{31.3} & \red{116.1} & \red{16.8e3} & \red{16.834} & \red{150.9} & \red{491.0} & \red{6.0e3} & \red{6.153e3}\\
\hline 
\end{tabular} 
\label{tbl:results} 
\end{table*}

\subsection{Expenses Analysis}
\begin{figure}[h]  
    \centering 
     \begin{subfigure}[b]{0.4\linewidth}
    \centering
        \begin{tikzpicture}[scale=0.4] 
          \begin{axis}[
                ybar,
                symbolic x coords={UG2D, P2D, P2PD, HED},
                xtick=data,
                ylabel={Average Expenses using Different Donation Algorithms (Euro)},
                ymin=0,
                bar width=15pt,
                enlarge x limits=0.4,
                nodes near coords,
                grid=major,
                ymode=log,
                ylabel style={ align=center, text width=7cm }, 
            ]
            \addplot[fill=tealshade] coordinates {(UG2D, 8588) (P2D, 7622) (P2PD, 1073333) (HED, 1081900)};
            \end{axis}
        \end{tikzpicture}
        \caption{ }
        \label{fig:expenses}
    \end{subfigure} \hfill
    \begin{subfigure}[b]{0.4\linewidth}
        \begin{tikzpicture}[scale=0.4]
          \begin{axis}[
                ybar,
                symbolic x coords={UG2D, P2D,  HED},
                xtick=data,
                ylabel={Average Cost using Different Donation Algorithms (Euro)},
                ymin=0,
                bar width=15pt,
                enlarge x limits=0.4,
                nodes near coords,
                grid=major,
                ymode=log,        
                ylabel style={ align=center, text width=7cm }, 
            ]
            \addplot[fill=tealshade] coordinates {(UG2D, 137.5) (P2D, 3.28) (HED, 133.41) };
            \end{axis}
        \end{tikzpicture}
        \caption{ }
        \label{fig:cost}
    \end{subfigure}
    \caption{\red{(a) Expenses of donated energy in Euro, and (b) Cost of externally donated energy per megawatt.}}
    \label{fig:expenses-cost-combined}
    
    \end{figure}
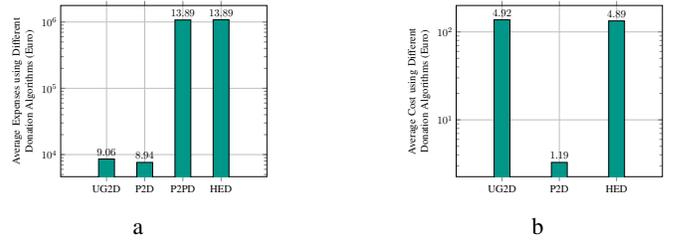

Figure \ref{fig:expenses} shows the average of donated energy expenses, where it includes the cost of the donated energy by external funds (where money transactions took place), and the price of donated energy by the peers (where there is no money transactions). 

Figure \ref{fig:cost} shows the average cost of externally donated energy that was paid for by external donors. The cost was calculated as the average of externally paid money divided by the average of externally donated energy. It is observed that P2D had the lowest cost, since the energy was sourced from peers only. Comparing HED with UG2D, it is noticeable that the cost of the donated energy was lower, because of having some of the energy sourced from the peers. We also note that the cost of donated energy in HED was higher than P2D. The reason is that most of the externally donated energy was sourced from the utility grid. 

\subsection{Quantitative Assessment of Donor and Recipient }
The average number of peers who sold energy to be donated was approximately \red{77}\% of the total peers in the P2D algorithm and \red{19.8}\% in the HED algorithm. Regarding the peers who donated energy, \red{97}\% of the peers donated energy in both the P2PD and HED algorithms. The percentages of donation receivers are \red{10}\%, \red{15.9}\%, \red{55.5}\%, and \red{56.0}\% in the UG2D, P2D, P2PD, and HED algorithms, respectively. 
\red{Finally, the simulation results indicate that 54\% of peer donors required energy at some point and subsequently became donees, when using P2PD. Similarly, in the case of HED, 54.5\% of donors also turned into donees.}



\section{Discussion} \label{discussion}

The P2PD and HED algorithms demonstrate a significant performance improvement of over 100\% in the total amount of donated energy\red{, in comparison to UG2D and P2D}. This outcome is expected, as these algorithms enable peers to contribute energy directly to others in need. However, this assumption may not hold in real-world applications. In scenarios where no peers are willing to donate, only the UG2D and P2D algorithms remain viable. Under these conditions, P2D shows a clear advantage, achieving an approximate \red{228}\% increase in donated energy compared to the UG2D algorithm. Furthermore, P2D aligns well with practical implementations, as it is likely to be adopted by prosumers due to its capability to offer more competitive energy pricing compared to the rates set by the utility grid. 

When comparing the HED and P2PD algorithms, the HED algorithm provides a slightly higher amount of donated energy than P2PD. Additionally, HED introduces a mechanism to reduce the load on peers by leveraging funds from external donors, thus enhancing the adaptability and feasibility of the donation process.

\section{Conclusions and Future Works} \label{conc}

In this study, we examined energy donations as a strategic approach to address energy poverty, presenting four distinct algorithms: \textit{Payment Donations to Utility Grid (UG2D)}, \textit{Payment Donations to Peers (P2D)}, \textit{Peer-to-Peer Donations (P2PD)}, and \textit{Hybrid Energy Donation (HED)}. Experimental results demonstrate the effectiveness of HED in increasing the aggregate amount of donated energy, identifying it as the most effective algorithm in mitigating energy poverty. In scenarios where peers are less inclined to donate energy, the P2D method outperforms UG2D, increasing the energy donation amount by \red{228}\%. Furthermore, we proposed ED-DAO, an approach leveraging the Distributed Autonomous Organization framework within the energy donation domain. ED-DAO facilitates a fully decentralized energy donation process while incorporating community governance.

Future work would focus on implementing a proof-of-concept for ED-DAO, with an emphasis on scalability and cost analysis. \red{This will serve as the first step toward developing a real-world model to be tested in practical scenarios. To facilitate such experimentation in a real-life context, it will be essential to examine the technological barriers, regulatory constraints, governance conflicts, and the economic and environmental impacts of ED-DAO}. Additionally, the integration of ENRD tokens to prioritize energy sales could provide an incentive mechanism, further motivating peers to participate in energy donations. \red{To further test the efficiencies of the proposed algorithm, experiments on other datasets, such as the dataset in \cite{zekiye2024blockchain}, are planned. Additionally, in this paper, the assumption is that all peers are willing to donate their excess energy; however, this may not always hold in real-world scenarios. Conducting more realistic simulations with varying percentages of peer donors would be valuable for a comprehensive analysis of the proposed donation algorithms.}

\section*{Acknowledgment}
This work is supported by TUBITAK (The Scientific and Technical Research Council of Türkiye) 2247-A National Outstanding Researchers Program Award 121C338;  ASPIRE and ZU ViP project EU2105; and ZU SIC Award 12092.

\bibliographystyle{IEEEtran}
\bibliography{IEEEabrv,references}

\end{document}